\documentclass[showpacs,superscriptaddress,preprintnumbers,amsmath,amssymb,twocolumn,prl]{revtex4}
\usepackage{amssymb,amsmath}
\usepackage{graphics}
\usepackage{epstopdf}
\usepackage{epsfig}
\usepackage{color}
\usepackage{amsfonts}
\usepackage{bm}
\usepackage{amscd}

\begin{document}

\title{Coherent control of artificial molecules using an Aharonov-Bohm magnetic flux}

\author{Matisse Wei-Yuan Tu}
\affiliation{Department of Physics, National Cheng Kung University,
Tainan 70101, Taiwan} \affiliation{Advanced Science Institute,
RIKEN, Saitama 351-0198, Japan}
\author{ Wei-Min Zhang}
\email{wzhang@mail.ncku.edu.tw} \affiliation{Department of Physics,
National Cheng Kung University, Tainan 70101, Taiwan}
\affiliation{Advanced Science Institute, RIKEN, Saitama 351-0198,
Japan}
\author{ Franco Nori}
\email{fnori@riken.jp} \affiliation{Advanced Science Institute,
RIKEN, Saitama 351-0198, Japan} \affiliation{Physics Department, The
University of Michigan, Ann Arbor, Michigan 48109-1040, USA.}

\begin{abstract}
Bonding and anti-bonding states of artificial molecules have been
realized in experiments by directly coupling two quantum dots.
Without a direct coupling between two nearby quantum dots, here we
show that a continuous crossover, from symmetric to anti-symmetric
molecular state, can be achieved by changing the flux through a
double quantum dot Aharonov-Bohm (AB) interferometer.  We explicitly
present the flux-dependent real-time processes of molecular-state
formation.  In contrast to the transport current, which has a $2\pi$
period, the quantum state of the DQD molecule has a $4\pi$ period in
the AB flux.
\end{abstract}

\date{\today}


\maketitle

It is important to tailor quantum states, especially, to control the
coherent phase between two superposition states.  In the past
decades, artificial atoms and molecules in solid-state systems, such
as superconducting Josephson junctions \cite{You011590} and
semiconductor quantum dots (QDs)
\cite{Hanson071217,Buluta11104401,Morton11345} have provided novel
platforms for exploring such quantum-coherent effects.
Due to the tunability of various electronic couplings, double
quantum dot (DQD) systems, which are archetypes of artificial
molecules, have attracted considerable attention.  Using
direct-tunnel coupling, the coherence of charge states have been
investigated with Aharonov-Bohm (AB) interferometers in recent
experiments \cite{Hatano11076801,Yamamoto12247}. However, the
couplings to the electron reservoirs (electrodes used for
measurements and controls), generally induces decoherence to the
quantum state of the DQD molecule.  Here, we show that for an
uncoupled DQD in an AB interferometer, such decoherence can be
suppressed with an asymmetric design of the device geometry.
Therefore, by solely tuning the AB flux, the coherent control of the
DQD molecule (from the symmetric to the anti-symmetric state) can be
realized. Furthermore, we find that the period of the quantum state
of the DQD in the AB flux is $4\pi$.  The transport current,
obtained by averaging the DQD states, possesses a period of $2\pi$.
The coherence of the DQD molecular state and the coherence of
electron transport therefore manifest themselves fundamentally
different through the AB flux.

Coupled DQDs have been theoretically proposed as qubits
\cite{Loss98120,Blick} and experimentally realized
\cite{Buluta11104401,Morton11345,Fujisawa06759,Gorman05090502,petta051280,Petersson10246804}.
The feasibility of realizing various one- and two-electron molecular
states with tunable tunneling and exchange couplings was
demonstrated \cite{Holleitner0270,Hatano05268}.  Furthermore, the
coherence of electron transport has been investigated with a single
QD in AB interferometers \cite{Yacoby954047}.
Combining an inter-dot tunnel coupling with a magnetic flux has also
been studied theoretically \cite{Loss001035,Kang04117,Kubo06205310}
and experimentally
\cite{Holleitner01256802,Sigrist06036804,Hatano11076801,Yamamoto12247}.
In particular, controlling the molecular-state through AB phases is
of recent experimental interest \cite{Hatano11076801}.  Although
tunneling to the electrodes may be turned off to avoid the
electron-reservoirs-induced decoherence, such tunnelings are
indispensable for the AB effect.  Thus controlling the molecular
coherence through the AB flux is a new challenge.

\begin{figure}[h]
\includegraphics[width=7cm]{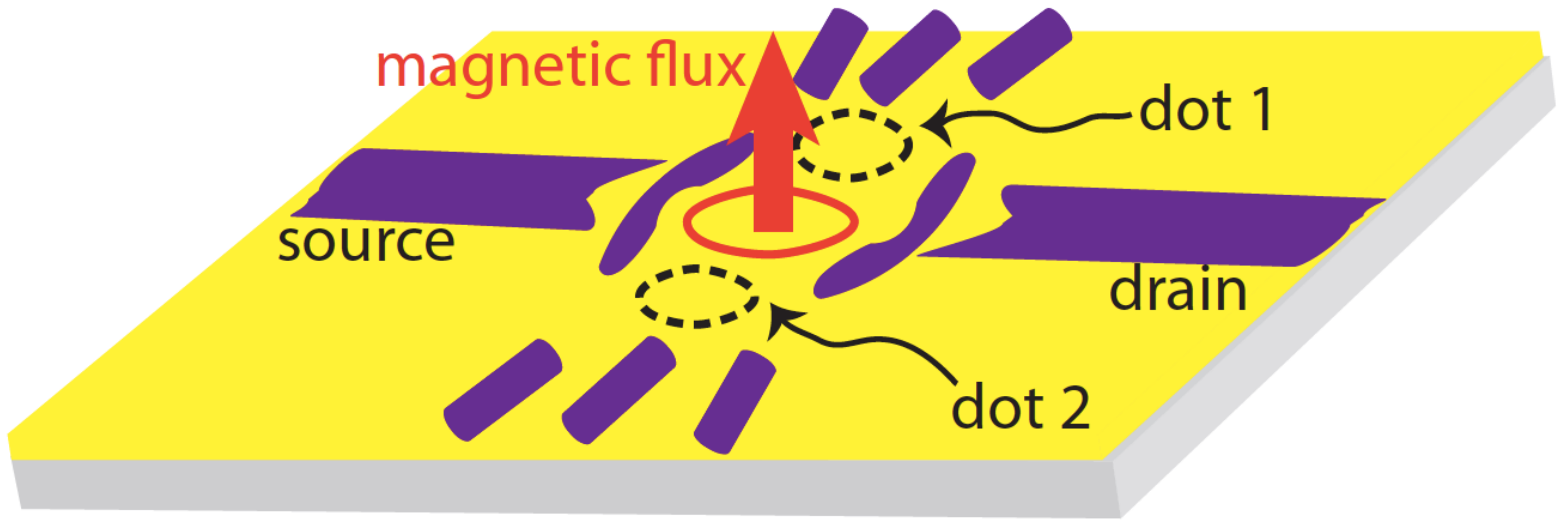}
\caption{(color online).  A schematic diagram of a pair of uncoupled
quantum dots in an Aharonov-Bohm interferometer.
 \label{fig1} }
\end{figure}

Here we consider uncoupled DQDs embedded in an AB interferometer, as
sketched in Fig.~\ref{fig1}.  In contrast to previous theoretical
studies, which focus on quantum transport
\cite{Kang04117,Kubo06205310}, here we directly exploit the quantum
state of the artificial molecule.  By explicitly analyzing the
decoherence through the AB flux, we deduce the proper geometry of
the DQD for coherent control over the molecular states. The
time-resolved formation processes of various molecular states,
determined by different AB fluxes, are explicitly presented.

\emph{The model system and its exact solution}.---To focus on the influence of the AB flux on
the quantum state of the artificial molecule, we consider only
polarized non-interacting electrons.  The total Hamiltonian of the system  is
conventionally \cite{Kubala02245301} given by ${\cal H}={\cal
H}_{\rm s}+{\cal H}_{\rm E}+{\cal H}_{\rm T}$, in which ${\cal H}_{\rm
s}=\sum_{i}E_{i}a^{\dag}_{i}a^{}_{i}$ describes an uncoupled DQD and
${\cal H}^{}_{\rm E}=\sum_{\alpha \bm{k}} \epsilon^{}_{\alpha
\bm{k}}c^{\dag}_{\alpha \bm{k}} c^{}_{\alpha \bm{k}}$ is the
Hamiltonian for the leads with $\alpha=L (R)$ labeling the source
(drain) lead, and $ {\cal H}_{\rm T}=\sum_{j
\alpha\bm{k}}[V_{j\alpha} c^{\dag}_{\alpha \bm{k}}a_{j}+{\rm H.c.}]
$ depicts the coupling between the central dot system
and the leads.  Here $a^{\dagger}_{i}$ ($a^{}_{i}$) and
$c^{\dag}_{\alpha \bm{k}}$ ($ c_{\alpha \bm{k}}$) are the electron
creation (annihilation) operators for the electronic levels $i$ and
$\bm k$ in the dot system and the lead $\alpha$, respectively. The
tunneling amplitudes harbor the applied magnetic flux $\Phi$ via
$V_{1L}^{*}=V_{2L}=|V_{L}|e^{i\phi/4}$, and
$V_{1R}=V_{2R}^{*}=|V_{R}|e^{i\phi/4}$, where
$\phi=2\pi\Phi/\Phi_{0}$ and $\Phi_{0}=h/e$ is the flux quantum. The
line-widths induced by tunneling are then given by
$\Gamma_{\alpha}=2\pi|V_{j\alpha}|^{2}\varrho_{\alpha}$, where
$\varrho_{\alpha}$ is the density of states in the lead $\alpha$.
The DQD molecular states are governed by the following master equation
\cite{Tu08235311}:
\begin{align}
&\frac{d}{dt}{\rho}(t)=-i[{\cal H}_{\rm
s},\rho(t)]+\sum_{i\alpha}[{\cal L}^{+}_{i\alpha}(t)+{\cal
L}^{-}_{i\alpha}(t)]\rho(t)  ,   \label{rho}
\end{align}
where ${\cal L}^{\pm}_{i\alpha}(t)$ are the
superoperators describing dissipations and fluctuations induced by
the tunnel coupling to the electrodes (for details, see Ref.
[\onlinecite{Tu08235311}]). Denoting the state of the empty DQD by
$|0\rangle$, one electron on the first and the second dot by
$|1\rangle$ and $|2\rangle$, respectively, and the state of both
dots occupied by $|3\rangle$, the density matrix $\rho(t)$ can be
generally expressed as
\begin{align}
\rho(t)=\begin{pmatrix} \rho^{}_{00}(t) & 0 & 0 & 0 \\ 0 &
\rho_{11}(t) & \rho_{12}(t) & 0 \\ 0 & \rho_{21}(t) & \rho_{22}(t) &
0
\\ 0 & 0 & 0 & \rho^{}_{33}(t)
\end{pmatrix} \  \label{srdm}
\end{align} where $\rho_{ij}=\langle i|\rho|j\rangle$ with
$i,j=0,1,2,3$.  The molecular state, featured as one electron in the
DQD shared between the two orbitals of the dots, is embedded in the
central $2\times 2$ block matrix of Eq.~(\ref{srdm}).  In particular, the
coherence between the two atomic orbitals of the DQD molecule is
characterized by the off-diagonal element, $\rho_{21}$.  To see how
molecular states in this DQD are formed in time, we solve the
master equation (\ref{rho}) with the initial preparation of  empty DQD,
namely, $\rho_{00}(0)=1$ and $\rho_{ij}(0)=0$, for all $i\ne0$,
$j\ne0$.  The explicit solution of each matrix element gives
\begin{align}
& \rho_{11}(t)=v_{11}(t) - \det \bm v(t), \nonumber \\
& \rho_{22}(t)=v_{22}(t) - \det \bm v(t), \nonumber \\
\rho_{12}&(t)=v_{12}(t), ~~ \rho_{21}(t)=v_{21}(t), \nonumber \\
\rho_{00}(t)&=\det[I-\bm v(t)], ~~\rho_{33}(t)=\det \bm v(t)
,\label{rho_v}
\end{align} with $I$ being an identity matrix and
\begin{align}
\bm v(t)\!\!=\!\! \int\!\! \frac{d\omega}{2\pi}{\bm u}
(t,\omega)\sum_\alpha
f_\alpha(\omega) \Gamma_{\alpha}\begin{pmatrix} 1 & e^{\pm i \phi/2}\\
e^{\mp i\phi/2} & 1 \end{pmatrix} \!\! {\bm u}^\dag(t,\omega)
\label{sol_v}
\end{align}is a $2\times2$ hermitian matrix, where $f_\alpha(\omega)$
is the Fermi distribution function of the reservoirs, the upper (lower) sign is
for $\alpha=L$ ($R$), and ${\bm u}
(t,\omega)=\int_{t_{0}}^{t}d\tau{e}^{i\omega(t-\tau)}{\bm u}(\tau)$
with
\begin{align}
\label{sol_u}
{\bm u}(\tau)=\exp\left[-\begin{pmatrix} iE_{1}+\Gamma & \Gamma_{c}(\phi) \\
\Gamma_{c}^{*}(\phi) & iE_{2}+\Gamma
\end{pmatrix}\tau \right].
\end{align} Here we have defined
$\Gamma_{c}(\phi)\!\!=\!\!\Gamma\cos(\phi/2)+i\delta\Gamma\sin(\phi/2)$
with $\Gamma\!\!=\!\!\Gamma_{L}\!+\!\Gamma_{R}$ and
$\delta\Gamma\!\!=\!\!\Gamma_{L}\!-\!\Gamma_{R}$.  The functions
$\bm u(t)$ and $\bm v(t)$ are indeed the retarded and correlation Green functions
in the Schwinger-Keldysh nonequilibrium Green function theory \cite{SK65}.
The AB flux $\phi$, the coupling asymmetry $\delta\Gamma$, the non-degeneracy
$\delta\!E\!\!=\!\!E_{1}-E_{2}$, and the nonequilibrium dynamics from
the electron tunnelings, all influence the consequent quantum states
of the DQD molecule.

\emph{Coherent phases controlled by the AB flux}.---
To have a clear picture of the coherence of the DQD molecular state, 
let us look at the off-diagonal
matrix element $\rho_{12}(t)$ in Eq.~(\ref{rho_v}) first, in the 
steady-state limit ($t\gg\Gamma^{-1}$).  The general solution gives
\begin{widetext}
\begin{align}
\label{rho21}
\rho_{21}=\frac{1}{2\pi}&\left[\tan^{-1}\left(\frac{eV}{2\Gamma_{+}(\phi)}\right)+\tan^{-1}\left(\frac{eV}{2\Gamma_{-}(\phi)}\right)\right]
\left[\frac{\delta\Gamma}{\Gamma}\cos\frac{\phi}{2}-i\sin\frac{\phi}{2}\right]+\frac{\delta\!E}{4\pi\gamma(\phi)}
\left[\frac{1}{\Gamma_{+}(\phi)}\tan^{-1}\left(\frac{eV}{2\Gamma_{+}(\phi)}\right)
\right. \nonumber\\&~~~~~-\left.
\frac{1}{\Gamma_{-}(\phi)}\tan^{-1}\left(\frac{eV}{2\Gamma_{-}(\phi)}\right)\right]
 \left\{ \frac{1}{\Gamma}\left[
(\Gamma^{2}-\delta\Gamma^{2})\sin\frac{\phi}{2}-\delta\Gamma\delta\!E\cos\frac{\phi}{2}
\right]-i\delta\!E\sin\frac{\phi}{2} \right\} ,
\end{align}
\end{widetext}
where $\gamma(\phi)$=$\sqrt{\Gamma^{2}\cos^{2}(\phi/2)
+\delta\Gamma^{2}\sin^{2}(\phi/2)-\delta\!E^{2}}$ and
$\Gamma_{\pm}(\phi)=2^{-1}(\Gamma\pm\gamma(\phi))$.  Here, we have
also applied a bias $\mu_{L}=eV/2=-\mu_{R}$ at zero temperature.  The
full complexity of decoherence is revealed through
Eq.~(\ref{rho21}). Due to the severe decoherence in such system, it
can be proven \cite{Tu11115318} that the coherent phase $\varphi$
(in the off-diagonal matrix element
$\rho_{21}=|\rho_{21}|e^{i\varphi}$) between the two atomic orbitals
can only take the values of 0, $\pm\pi/2$ or $\pi$ for arbitrary
flux. This applies for the often-used condition of degeneracy
$\delta\!E=0$ and symmetric coupling $\delta\Gamma=0$.  The
decoherence-induced discretization of the coherent phase hinders the
manipulation of the coherent phase of molecular states.  However,
when the DQD is non-degenerate and couples asymmetrically to the
left and the right leads ($\delta\Gamma\ne0$), we find that the
coherent phase $\varphi$ can be continuously tuned by the AB flux.

\begin{figure}[h]
\includegraphics[width=0.45\textwidth]{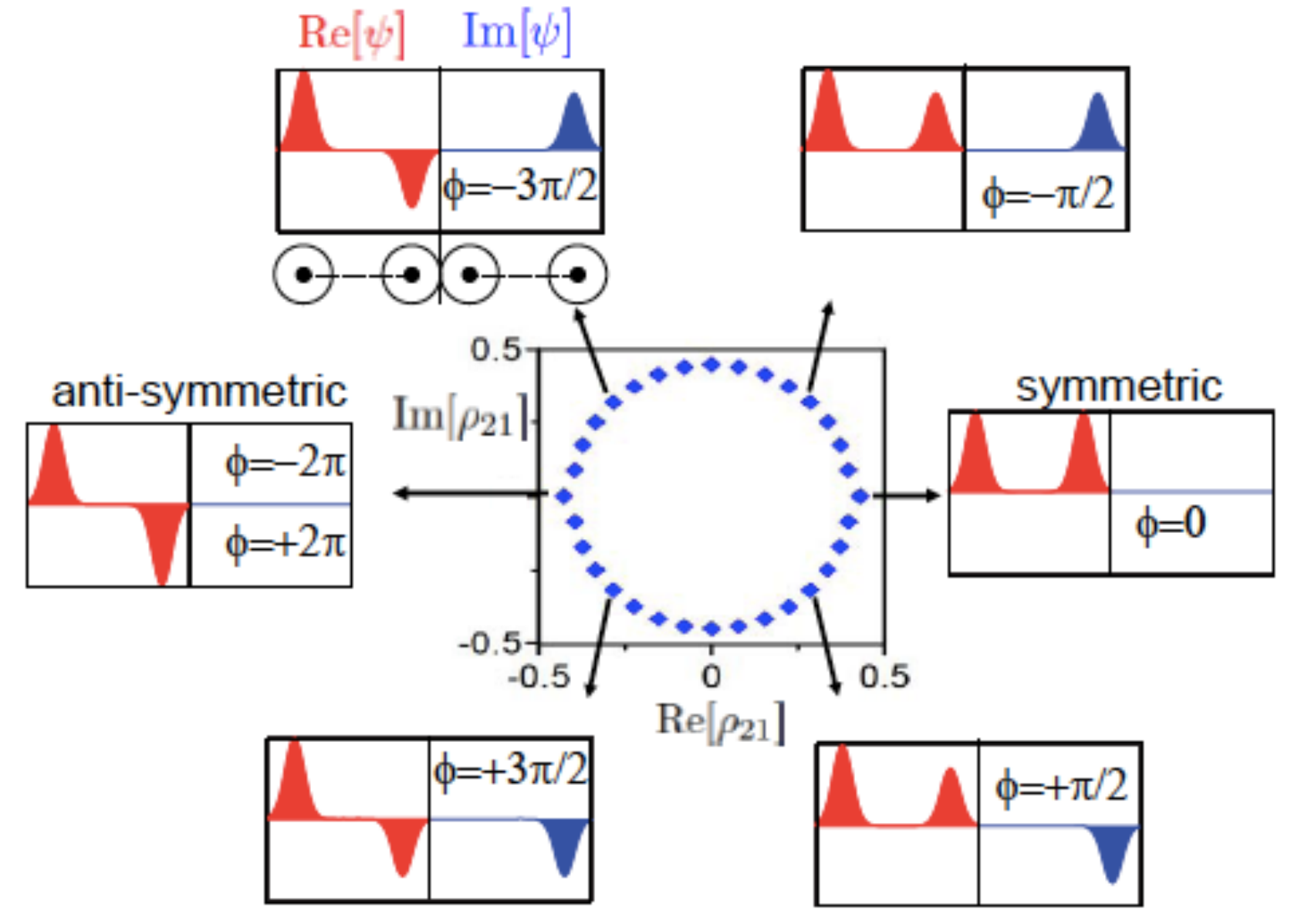}
\caption{(color online).  Control of the coherent phase of the DQD
molecule by the AB flux.  The explicit solution of $\rho_{21}$ in
the steady-state limit is shown by the ``blue diamonds" on the
central panel.  Each diamond corresponds to an AB flux value, taken
from $\phi=0$ to $\phi=\pm2\pi$ with $\pi/8$ steps.  The
wavefunctions on the DQD molecules are illustrated for various
values of the AB flux.  A DQD is indicated by two circles with
centers connected by a dashed line (no inter-dot coupling) below the
diagrams for $\phi=-3\pi/2$.  Both the real (red) and the imaginary
(blue) parts are shown, so one sees how the AB flux changes the
coherent phase between the atomic orbitals.  Other parameters are
$\delta\!E=0$, $eV=6\Gamma$ at $k_{B}T=\Gamma/20$, which 
are also used in the following figures, unless specified.
 \label{fig2} }
\end{figure}

In order to achieve typical molecular states,
the DQD is set at degeneracy ($\delta\!E=0$).  The second term in
Eq.~(\ref{rho21}) vanishes. Equations (\ref{rho_v},\ref{rho21}) show
that the formation of molecular states is essentially determined by
the applied bias and the coupling asymmetry to the source
and the drain.  The basic setup of zero bias (which is used for
examining quantum transport) leads to $\rho_{21}=0$ (because the DQD
is in equilibration with the reservoirs) and is not interested here.
With a large bias, $eV\gg\Gamma$, we find
\begin{align}
&\rho_{21}= \left\{\begin{array}{ll}\frac{1}{2}\big[(\delta\Gamma/\Gamma)
\cos\frac{\phi}{2}-i\sin\frac{\phi}{2} \big] & \mbox{if $\phi\ne0$}, \\ ~ & ~ \\
\frac{1}{4}(1+\delta\Gamma/\Gamma) & \mbox{if $\phi=0$.}
\end{array} \right. \label{rho12_evlarge}
\end{align}
Equation~(\ref{rho12_evlarge}) clearly shows the
controllability of the coherent phase between the two atomic
orbitals of the DQD molecule through the AB flux.  It also explicitly
reveals the necessity of the asymmetry in couplings,
$\delta\Gamma\ne0$.  In the case of
symmetric coupling, $\delta\Gamma=0$,
Eq.~(\ref{rho12_evlarge}) shows that the real part vanishes for $\phi \neq 0$
so that the coherent phase $\varphi$ is localized at $\pi/2$,
except for $\phi=0$ where the coherence phase is
restricted to $0$, as it has been pointed out in \cite{Tu11115318}.
With the larger asymmetry $\delta\Gamma$, the coherence
amplitude $|\rho_{21}|$ linearly increases  and the coherence phase is
continually driven by the AB flux, as seen from Eq.~(\ref{rho12_evlarge})
\cite{expn2}.
 By setting $\delta\Gamma\lesssim\Gamma$, we obtain
$\rho\approx|\psi(\phi)\rangle\langle\psi(\phi)|$, where
\begin{align}
\label{state_phi}
|\psi(\phi)\rangle=\frac{1}{\sqrt{2}}\big[|1\rangle+\exp(-i\phi/2)|2\rangle
\big].
\end{align}  A continuous transition from the symmetric,
$|\psi(0)\rangle=\left(|1\rangle+|2\rangle\right)/\sqrt{2}$, to the
anti-symmetric state,
$|\psi(\pm2\pi)\rangle=\left(|1\rangle-|2\rangle\right)/\sqrt{2}$,
 is achieved by changing the AB flux, as shown in
Fig.~\ref{fig2} (see captions).  Interestingly we find that the
period of the state of the DQD molecule is $4\pi$, rather than
$2\pi$ in the AB flux as one expected.

\emph{Real-time processes of molecular-state formations}.---
The full information of the quantum state of the DQD molecule at
finite temperature is depicted by the time-dependent reduced 
density matrix.   We can write the central block matrix of Eq.~(\ref{srdm}) as
\begin{align} \label{qdm}
\rho_{\rm q}(t)=\frac{1}{2}\big[I + {\bm r}(t)\cdot {\bm
\sigma}\big]- \frac{1}{2}\big[\rho_{00}(t)+\rho_{33}(t)\big]I \ ,
\end{align}
where $\bm \sigma=(\sigma_{x},\sigma_{y},\sigma_{z})$ consists of
the Pauli matrices and ${\bm r}(t)=2\{{\rm Re}\rho_{21}(t), {\rm
Im}\rho_{21}(t), \rho_{11}(t)-\rho_{22}(t)\}$, is the polarization
vector for the molecular states.  So the dynamics of molecular-state
formations can be visualized through the motion of the polarization
vector with the Bloch sphere.  Also, the leakage out of the
one-electron state-space can be easily seen from the term proportional to
the probability of the empty and the double occupied states,
$\rho_{00}(t)+\rho_{33}(t)$.

\begin{figure*}[tbp]
\begin{center}
\includegraphics[width=0.7\textwidth]{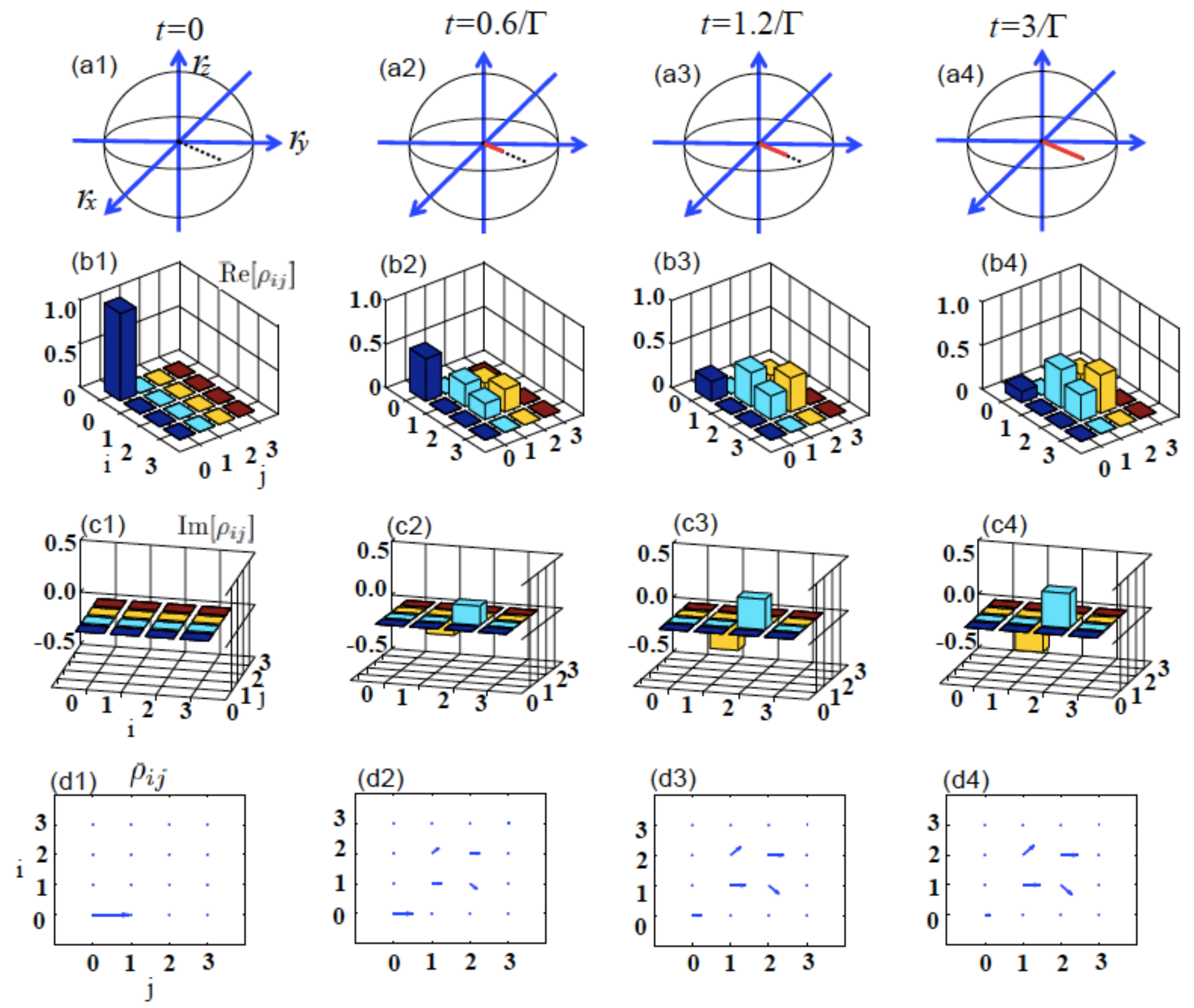}
\caption{(color online).  Typical process for forming molecular
states.  The dashed black line in (a1) to (a4) is the trajectory
taken by ${\bm r}(t)$ from $t=0$ in (a1), starting from the origin,
to $t=3/\Gamma$ in (a4), where it almost touches the surface of the
sphere.  The red strip in each plot is the trajectory up to the
corresponding time points, as shown above the spheres.  From the
trajectory, we see the coherent phase $\varphi$ (which is the angle
made by ${\bm r}(t)$ with $r_{x}$ axis) has been fixed after the
electron is injected into the DQD.  Plots (b1) to (b4) display the
real part of the reduced density matrix of the DQD system, while the
imaginary part is plotted in (c1) through (c4).  The coherent phase
$\varphi$ between the two atomic orbitals is better visualized
through the vector plots (d1) to (d4).  Every arrow represents an
element of the reduced density matrix $\rho_{ij}$, whose horizontal
projection stands for the real part and the vertical projection
stands for the imaginary part.  The AB flux here is $\phi=-\pi/2$.
\label{fig3} }
\end{center}
\end{figure*}

 In Fig.~\ref{fig3},
we plot the evolution of the full reduced density matrix of the DQD
molecule.  Initially, the DQD is prepared in
an empty state, $\rho_{00}(0)=1$ as shown by
Fig.~\ref{fig3}(b1,c1,d1) and ${\bm r}(0)=0$ given in
Fig.~\ref{fig3}(a1) (where the length of the red strip is zero).
After injecting electrons from the left and the right reservoirs,
$\rho_{00}$ decreases [see Fig.~\ref{fig3}(b1) to (b3)] while the
electron occupation and coherence increase with time [see plots (b1)
to (b3), (c1) to (c3), and also (d1) to (d3) in Fig.~\ref{fig3}].
The coherent phase $\varphi$ between the atomic orbitals has been
fixed shortly after the electron injection [see Fig.~\ref{fig3}(a1)
to (a4) and also (d2) to (d4)].  Then $|{\bm r}(t)|$ grows in time
with fixed $\varphi$, and finally a stable molecular state,
$\rho\approx|\psi\rangle\langle\psi|$, where
$|\psi\rangle=\left(|1\rangle+e^{-i\phi/2}|2\rangle\right)/\sqrt{2}$,
is reached in a short time, of about $3\Gamma^{-1}$.  Note that due
to possible leakage, see Fig.~\ref{fig3}(b3) where $\rho_{00}$ has a
small finite value, the DQD is not in a perfect pure state.  But the
situation can be optimized by changing the bias and the coupling
asymmetry, as shown by Eq.~(\ref{rho12_evlarge}).

\begin{figure}[tbp]
\includegraphics[width=0.4\textwidth]{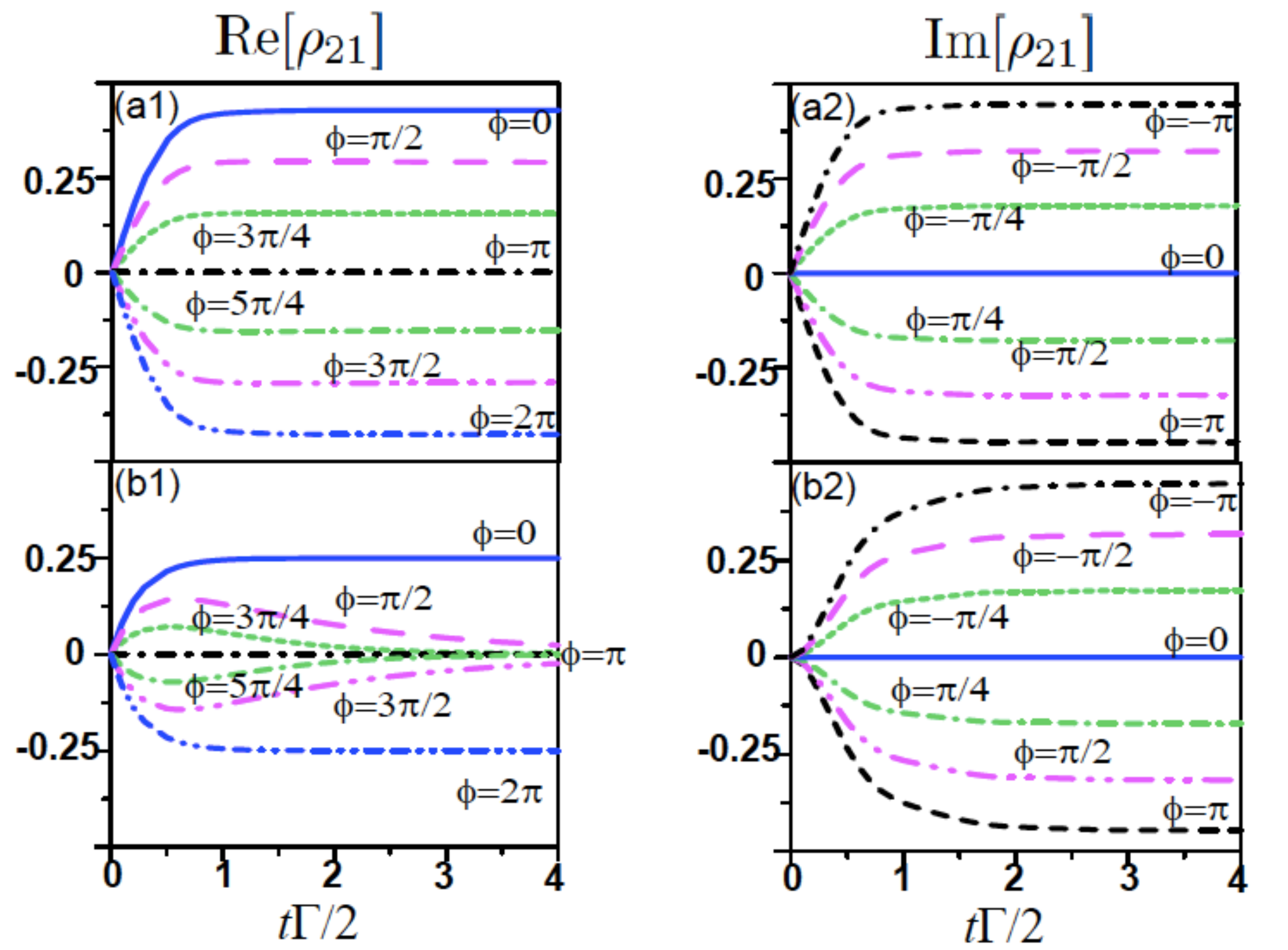}
\caption{(color online).  The time evolutions of $\rho_{21}$.  The
plots (a1,a2) for $\delta\Gamma=0.9\Gamma$ and (b1,b2) for
$\delta\Gamma=0$.  Also (a1,b1) give ${\rm Re}\rho_{21}$, and (a2,b2) give ${\rm
Im}\rho_{21}$.
 \label{fig4} }
\end{figure}

To
better understand the role played by the AB flux, we show the time
evolutions of $\rho_{21}$ in Fig.~\ref{fig4} under various values of
$\phi$. Figure~\ref{fig4}(a1,a2) show the process of coherence
generation for a strong asymmetric coupling (with $\delta\Gamma
=0.9\Gamma$).  The rate of approaching steady-coherent-molecular
states is only weakly dependent on the flux. The stable molecular
states are soon reached after a few $\Gamma^{-1}$. This is totally different
from the symmetric coupling ($\delta\Gamma=0$),
see Fig.~\ref{fig4}(b1,b2). It  shows the flux-dependent decays
of ${\rm Re}\rho_{21}$,  due to the
severe decoherence in the symmetric coupling.
Therefore, the coupling asymmetry can strongly suppress the
decoherence induced by electron tunnelings, and make the coherence
control of the QDQ molecule feasible.

\emph{Discussions}.---  The general solution 
shows that the quantum state of the DQD molecule has a period of $4\pi$
in the AB flux.  It is an intrinsic property of this pseudo-spin
system, independent of the coupling geometry and the bias configurations.
Besides, we have also calculated the tunneling current to reservoir
$\alpha=L,R$ within the same framework \cite{Tu08235311} with the result
$I_{\alpha}(t)=e~\sum_{i}{\rm tr}_{\rm s}[{\cal
L}^{+}_{i\alpha}(t)\rho(t)]$.
The steady-state transport current $I=\frac{1}{2}(I_{L}-I_{R})$
is then given by
\begin{align}
I(\phi)=\int\frac{d\omega}{2\pi}[f_{L}(\omega)-f_{R}(\omega)]
\mathcal{T}(\omega,\phi), \label{sscurrent}
\end{align} where the transmission coefficient is given by
\begin{align}
\mathcal{T}(\omega,\phi)=\frac{(\Gamma^{2}-\delta\Gamma^{2})[\omega^{2}
\cos^{2}\frac{\phi}{2}+\frac{1}{4}\delta\!E\sin^2\frac{\phi}{2}]}
{[\omega^{2}+\Gamma_{+}^{2}(\phi)][\omega^{2}+\Gamma_{-}^{2}(\phi)]}.
\label{ctm}
\end{align}
By taking $\delta\Gamma=0$ it reproduces the result in
Ref.~\cite{Kubala02245301}.  Equation (\ref{ctm}) clearly shows
that the transport current has a period in the AB flux of $2\pi$.
This $2\pi$ period, as a feature for the coherence of transport, is
well known and has been observed in experiments
\cite{Yacoby954047, Holleitner01256802, Sigrist06036804}.  The $4\pi$ period,
a nontrivial character of the quantum state of the DQD molecule, requires
further experimental investigation.  Note that although the coherent phase
of the off-diagonal density matrix element is gauge-dependent, the AB flux
dependence of the coherence phase and its periodicity are both
independent of the gauge choice.

In summary, we have demonstrated the effectiveness of the AB flux 
for the coherence control of DQD
artificial molecules.  We have analyzed the AB flux-dependent
conference controlling, through the asymmetric coupling of the DQD
to the electron reservoirs.  When a large bias is applied with a strong asymmetry in
couplings to the source and the drain, coherent control by the AB
flux can be easily achieved.  The decoherence induced by the electron
tunnelings can be efficiently suppressed. We also find that the period
of the quantum state of the DQD molecule in the AB flux is $4\pi$.
The revelation of the underlying quantum-coherence of the molecular
states is thus beyond the usual transport measurement. The
verifications of these molecular states would rely on a suitable
quantum-state-tomography protocol for further investigations.  We hope
that this theory for artificial molecules could inspire new experiments
on coherence control of molecular states via the AB flux, and
become also useful for the quantum emulation \cite{Buluta09108}
of artificial molecular processes.

\begin{acknowledgements}
This work is supported in part by the National Science Council of
ROC under Contract No. NSC-99-2112-M-006-008-MY3 and  we acknowledge
support of computing facility from HPC center of national Cheng Kung
university and National Center for Theoretical Science.  FN is
partially supported by the ARO, NSF grant No. 0726909, JSPS-RFBR
contract No. 12-02-92100, Grant-in-Aid for Scientific Research (S),
MEXT Kakenhi on Quantum Cybernetics, and the JSPS via its FIRST
program.
\end{acknowledgements}


\vspace*{-0.2 in}

\begin{thebibliography}{999}

\bibitem{You011590}J. Q. You and F. Nori, Nature \textbf{474}, 590
(2011); Phys. Today \textbf{58}, 42 (2005).

\bibitem{Hanson071217} R. Hanson, L. P. Kouwenhoven, J. R. Petta, S. Tarucha,
and, L. M. K. Vandersypen, Rev. Mod. Phys. \textbf{79}, 1217 (2007).

\bibitem{Buluta11104401}I. Buluta, S. Ashhab and F. Nori, Rep. Prog. Phys.,
\textbf{74}, 104401 (2011).

\bibitem{Morton11345}J. J. L. Morton,   D. R. McCamey,    M. A.
Eriksson, and S. A. Lyon, Nature \textbf{479}, 345 (2011).

\bibitem{Hatano11076801}T. Hatano, T. Kubo, Y. Tokura, S. Amaha, S. Teraoka, and S. Tarucha, Phys. Rev. Lett. \textbf{106}, 076801 (2011).

\bibitem{Yamamoto12247}M. Yamamoto, S. Takada, C. Bauerle, K. Watanabe, A. D. Wieck and S. Tarucha,  Nat. Nanotech. \textbf{7}, 247 (2012).

\bibitem{Loss98120}D. Loss and D. P. DiVincenzo, Phys. Rev. A \textbf{57}, 120
(1998).

\bibitem{Blick}R. H. Blick and H. Lorenz, in \textit{Proceedings of the IEEE International
Symposium on Circuits and Systems}, edited by J. Calder (IEEE,
Piscataway, NJ, 2000), Vol. II, p. 245.


\bibitem{Fujisawa06759}T. Hayashi, T. Fujisawa, H. D. Cheong, Y. H. Jeong, and Y.
Hirayama, Phys. Rev. Lett. \textbf{91}, 226804 (2003);

\bibitem{Gorman05090502}J. Gorman, D. G. Hasko, and D. A. Williams,
Phys. Rev. Lett. \textbf{95}, 090502 (2005).

\bibitem{petta051280}J. R. Petta, A. C. Johnson, J. M. Taylor, E. A. Laird,
A. Yacoby, M. D. Lukin, C. M. Marcus, M. P. Hanson, A. C. Gossard.
Science \textbf{309}, 1280 (2005).

\bibitem{Petersson10246804}K. D. Petersson, J. R. Petta,  H. Lu, and A. C.
Gossard, Phys. Rev. Lett., \textbf{105}, 246804 (2010).

\bibitem{Holleitner0270}A. W. Holleitner, R.H. Blick, A.K. Huttel, K. Eberl, and J. P. Kotthaus, Science \textbf{297}, 70 (2002).

\bibitem{Hatano05268}T. Hatano, M. Stopa, and S. Tarucha, Science \textbf{309}, 268
(2005).

\bibitem{Yacoby954047}A. Yacoby, M. Heiblum, D. Mahalu, and H. Shtrikman, Phys. Rev. Lett.
\textbf{74}, 4047 (1995).

\bibitem{Loss001035}D. Loss and E.V. Sukhorukov, Phys. Rev. Lett. \textbf{84}, 1035 (2000).

\bibitem{Kang04117}K. Kang and S. Y. Cho, J. Phys. Condens. Matter \textbf{16}, 117 (2004).

\bibitem{Kubo06205310}T. Kubo, Y. Tokura, T. Hatano, and S. Tarucha, Phys. Rev. B \textbf{74},
205310 (2006).

\bibitem{Holleitner01256802}
A. W. Holleitner, C. R. Decker, H. Qin, K. Eberl, and R. H. Blick,
Phys. Rev. Lett. \textbf{87}, 256802 (2001).

\bibitem{Sigrist06036804}
M. Sigrist, T. Ihn, K. Ensslin, D. Loss, M. Reinwald, and W.
Wegscheider, Phys. Rev. Lett. \textbf{96}, 036804 (2006).

\bibitem{Kubala02245301}B. Kubala and J. K\"{o}nig, Phys. Rev. B \textbf{65},
245301 (2002).

\bibitem{Tu08235311} M. W. Y. Tu and W. M. Zhang, Phys. Rev. B {\bf
78}, 235311 (2008); J. S. Jin, M. W. Y. Tu, W. M. Zhang, and Y. J.
Yan, New J. Phys. {\bf 12}, 083013 (2010).

\bibitem{SK65} J. Schwinger, J. Math. Phys. \textbf{2}, 407 (1961);
L. V. Keldysh, Sov. Phys. JETP, \textbf{20}, 1018 (1965).

\bibitem{Tu11115318} M. W. Y. Tu, W. M. Zhang, and J. S. Jin,
Phys. Rev. B \textbf{83}, 115318 (2011).

\bibitem{expn2} When $\phi=0$, the amplitude enhancement of the coherence is determined
by the sign of $\delta\Gamma$. When $\mu_{L}>\mu_{R}$, the asymmetry
of $\Gamma_{L}>\Gamma_{R}$ is preferred for larger $|\rho_{21}|$.

\bibitem{Buluta09108}I. Buluta and F. Nori, Science \textbf{326}, 108 (2009).

\end{thebibliography}
\end{document}